\documentclass[twocolumn,prl,showpacs,superscriptaddress]{revtex4}

\usepackage{graphicx}
\usepackage{rotating}
\usepackage{times}
\usepackage{amsmath}
\usepackage{amsfonts}
\usepackage{amssymb}
\usepackage{enumerate}
\usepackage{longtable}
\usepackage{dcolumn}
\usepackage{bm}
\usepackage[colorlinks,bookmarks=false,citecolor=blue,linkcolor=blue,urlcolor=blue]{hyperref}
\usepackage{color}
\begin{document}
\newcommand{\cred}{\color{red}}
\title{Supersolid phase induced by correlated hopping in spin-1/2 frustrated quantum magnets}

\author{K.P.~Schmidt}
\email{kaiphillip.schmidt@epfl.ch}
\affiliation{Institute of Theoretical Physics, \'{E}cole Polytechnique F\'{e}d\'{e}rale de Lausanne, CH 1015 Lausanne, Switzerland}
\author{J. Dorier}
\affiliation{Institute of Theoretical Physics, \'{E}cole Polytechnique F\'{e}d\'{e}rale de Lausanne, CH 1015 Lausanne, Switzerland}
\author{A.M.~L\"auchli}
\affiliation{Institut Romand de Recherche Num\'erique en Physique des Mat\'eriaux (IRRMA), CH-1015 Lausanne, Switzerland}
\author{F.~Mila}
\affiliation{Institute of Theoretical Physics, \'{E}cole Polytechnique F\'{e}d\'{e}rale de Lausanne, CH 1015 Lausanne, Switzerland}
\date{\rm\today}

\begin{abstract}
We show that correlated hopping of triplets, which is often the dominant source of kinetic
energy in dimer-based frustrated quantum magnets, produces a remarkably strong tendency to
form supersolid phases in a magnetic field. These phases are characterized by simultaneous
modulation and ordering of the longitudinal and transverse magnetization respectively. 
Using Quantum Monte Carlo and a semiclassical approach for
an effective hard-core boson model with nearest-neighbor repulsion on a square lattice, 
we prove in particular that a supersolid phase can exist even if the repulsion is not
strong enough to stabilize an insulating phase at half-filling. Experimental implications
for frustrated quantum antiferromagnets in a magnetic field at zero and finite temperature 
are discussed.  
\end{abstract}

\pacs{05.30.Jp, 03.75.Kk, 03.75.Lm, 03.75.Hh}

\maketitle
\begin{figure}
   \begin{center}
   \includegraphics[width=.9\columnwidth]{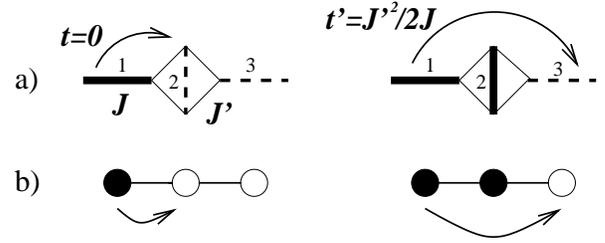}
   \end{center}
    \caption{Difference between single and correlated triplet hopping in a frustrated geometry.
     a) Spin language: thick solid (dashed) lines stand for dimer triplet (singlet), thin solid lines for inter-dimer coupling.
    b) Bosonic language: filled (open) circles denote hardcore boson sites which are occupied (empty). }
    \label{fig:hopping}
\end{figure}

The identification of exotic states of quantum matter in microscopic models is an important issue in
current research on strongly correlated quantum systems. The existence of a supersolid (SS) which
simultaneously displays crystalline order (solid) and long-range phase coherence
(superfluid, SF) has been definitely established recently thanks to extensive Quantum Monte Carlo (QMC) simulations
of a hard-core boson model with nearest-neighbor (n.n.) repulsion on the triangular lattice in the
context of cold atoms loaded into optical lattices.~\cite{wesse05,heida05,melko05}
The possibility to realize a supersolid phase in dimer-based quantum magnets, first pointed out 
by Momoi and Totsuka in
the context of SrCu$_2$(BO$_3$)$_2$~\cite{momoi00}, has been further
investigated very recently~\cite{ng06,sengu07,laflo07}. It relies on
the description of a polarized triplet on a dimer as a hard-core boson, a
convenient language we will mostly use throughout this paper. Some
trends have already emerged. In particular, for a supersolid to be realized,
triplets induced by the magnetic field should have a small kinetic energy as
compared to their mutual repulsion. In SU(2) models, this can be achieved if
the inter-dimer coupling is frustrated, as noticed in the context of
magnetization plateaux early on~\cite{mila}, and pointed
out recently for supersolids by Sengupta and Batista~\cite{sengu07}. This is not the whole story however.
First of all, even if the kinetic energy is small, the transition between insulating and superfluid phases can
be first order, thus preempting the possibility of a supersolid phase. 
This is for instance the case of hard-core bosons with n.n. repulsion on the
square lattice~\cite{batro00}, in contrast to the results of
Refs.~\cite{wesse05,heida05,melko05} on the triangular lattice.
Besides, when the hopping of a single triplet is strongly suppressed by frustration, the kinetic energy is not always suppressed accordingly. Indeed, correlated hopping terms much larger than single particle hopping
are often generated. To see this, suppose for instance that dimers are
locally arranged as in Fig.~\ref{fig:hopping} with $J'\ll J$, a situation
typical of frustrated magnets. The amplitude for a triplet to hop from dimer
1 to 2 is strictly equal to 0, while the amplitude to hop from 1 to 3 is
equal to $J'^2/2J$ if dimer 2 is occupied by a triplet, as first pointed out
in Ref.~\cite{momoi00_2}. In two-dimensional arrangements of frustrated coupled dimers, single particle hopping does not strictly vanish, but only appears at much higher order in perturbation (6th order for the Shastry-Sutherland model
realized in SrCu$_2$(BO$_3$)$_2$), while correlated hopping is typically a second order process~\cite{shinreview}.
So the problem of hard-core bosons with correlated hopping calls for further
investigation. 

In this Letter, we show that supersolidity is strongly favored in dimer-based frustrated magnets {\it because} correlated hopping is the principal source of kinetic energy. The bottom line of our analysis can be
summarized as follows: Under the effect of correlated hopping, a triplet cannot delocalize if it is alone, but it can in a crystalline arrangement of triplets with the appropriate geometry.
In particular, if an insulating phase with a geometry compatible with correlated hopping is realized at commensurate filling,
one may expect that upon adding particles, the crystalline order will be retained, leading to a supersolid phase.  As we shall see, this simple mechanism gives rise to a tendency toward supersolidity which is much stronger than anticipated, in fact so strong that a supersolid phase can exist even if the corresponding solid phase cannot be stabilized. 

These conclusions are based on an extensive investigation with Stochastic
Series Expansion (SSE) \cite{sandvik} Quantum Monte Carlo (QMC) simulations
and with a semiclassical approximation (SCA) of a minimal model of hard-core bosons on the square lattice defined by the Hamiltonian:
\begin{eqnarray}
 H&=&-t\sum_{\langle i,j\rangle} \left( b^\dagger_i b^{\phantom{\dagger}}_j+{\text{h.c.}}\right)+V\sum_{\langle i,j\rangle} n_i n_j -\mu\sum_i n_i\nonumber\\
&&-t^\prime\sum_i\sum_{\delta=\pm x;\delta^\prime=\pm y} n_i\left[b^\dagger_{i+\delta}b^{\phantom{\dagger}}_{i+\delta^\prime}+{\it h.c.}\right]
\end{eqnarray}
where $n_i=b^\dagger_i b^{\phantom{\dagger}}_i\in\{0,1\}$ is the boson
density at site $i$, $\mu$ the chemical potential, $t$ the n.n. hopping
amplitude, $t^\prime$ the amplitude of the correlated hopping, and $V$ the 
n.n. repulsion. 
The correlated hopping term describes a process where a particle
can hop along the diagonal of a square plaquette provided there is a particle on one of the other two sites
of the plaquette. In the context of weakly coupled dimers with intra and
inter-dimers exchange $J$ and $J'$, $t$ can be arbitrarily small, $V$ is of
order $J'$, $t'$ is of order $J'^2/J$, and $\mu=g\mu_B H - J +O(J')$\cite{momoi00}.
Throughout the paper, the energy scale is fixed by $t+t'=1$.

The case without correlated hopping ($t'=0$) has already been investigated thoroughly~\cite{schmid02}. For strong enough repulsion, an insulating phase with checkerboard (CB) order appears at half-filling. 
The phase diagram is symmetric about $n=1/2$ in that case due to particle-hole symmetry, and
the transition from the solid to the superfluid phase is first order with a jump in the
density~\cite{batro00}.
So, there is no supersolid phase in the absence of correlated hopping.
In the following, we study how this picture is modified when correlated hopping is introduced.
\begin{figure}
    \begin{center}
    \includegraphics[width=.9\columnwidth]{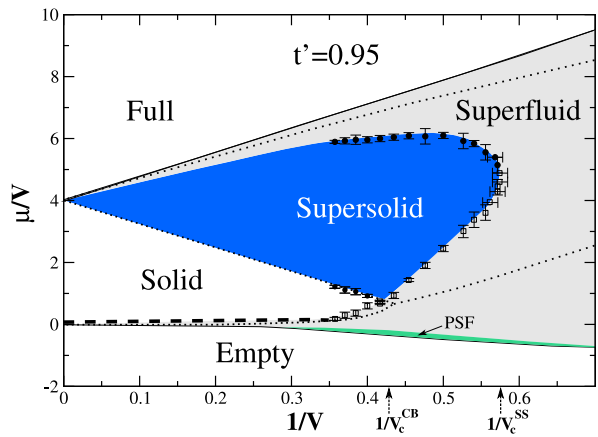}
    \end{center}
    \caption{(Color online) Zero-temperature phase diagram for $t'=0.95$ as a function of the $1/V$ versus $\mu/V$. 
     Open squares (closed circles) denote first (second) order phase transitions deduced from QMC data. Thin dotted lines are mean-field
results. The other lines (solid resp. thick dashed for second and first
     order transitions) are obtained by interpolating between numerical
     data. The data for low densities including 
the paired superfluid (PSF) have been taken from Ref.~\onlinecite{schmi06}. }
    \label{fig:pdV}
\end{figure}

Let us first briefly review some technical points. In SSE QMC, the various bosonic phases can be determined by studying the density $n$, the static structure factor at the wavevector $(\pi,\pi)$ relevant for the checkerboard solid
\begin{equation}
 S(\pi,\pi )=\frac{1}{N}\sum_{i,j} (-1)^{i-j} \langle n_i n_j\rangle\ ,
\end{equation}
and the superfluid stiffness
\begin{equation}
 \rho_{\rm S}=\frac{1}{2\beta L^2} \langle W_x^2 + W_y^2 \rangle,
\end{equation}
which signals the presence of a SF phase. Here $W_x$ and $W_y$ are the total winding numbers in $x$ and $y$ directions. If both order parameters $S(\pi,\pi)/N$ and $\rho_{\rm S}$ are finite in the thermodynamic limit, the system is in a SS phase. 

The SCA~\cite{scalet95} relies on a mapping onto a $S=1/2$ spin model using
the Matsubara-Matsuda~\cite{matsubara56} representation of
hard-core bosonic operators $b^\dagger=S^+$, $b =S^-$, and $S^z=1/2-n$ .
Note that the correlated hopping term transforms into a three-spin term. 
The energy, including zero-point fluctuations around the possibly
non-collinear classical ground state~\cite{note} calculated within linear spin-wave theory, is then minimized
assuming a 4-site unit cell to allow for broken translational symmetry. 
The superfluid phase corresponds to a ferromagnetic state with a non-zero component in the $xy$-plane, the
checkerboard solid phase to N\'eel order with wave-vector $(\pi,\pi)$ and with all spins parallel to the $z$ direction, while
the supersolid phase discussed below is close to N\'eel order, but the spins acquire a small ferromagnetically ordered non-zero component 
in the $xy$-plane. As a thumb rule, the approximation is expected
to be qualitatively and semi-quantitatively reliable provided the
mean-value of the local number of Holstein-Primakoff bosons is small
compared to 2S, which is the case here.

\begin{figure}
    \begin{center}
        \includegraphics[width=.9\columnwidth]{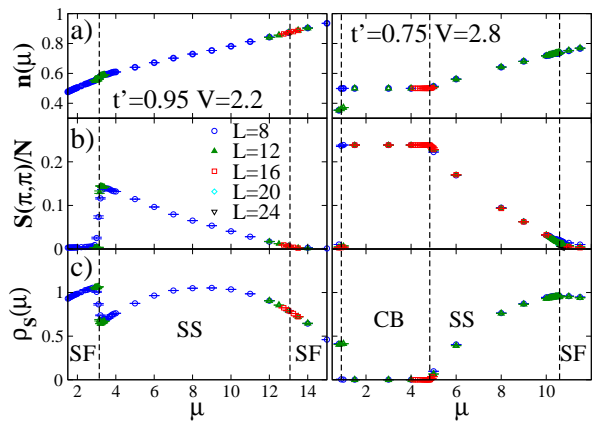}
    \end{center}
        \caption{(Color online) Bosonic phases revealed by: a) the density $n$, b) the static structure factor $S(\pi,\pi)/N$,
        and c) the superfluid stiffness $\rho$ as a function of the chemical potential for two representative cases. 
Left panel: $t'=0.95$ and $V=2.2$ ($1/V=0.455$). Right panel:
$t'=0.75$ and $V=2.8$ ($1/V=0.36$)
        SF stands for superfluid, CB for checkerboard solid, and SS for supersolid.}
    \label{fig:SS_T0}
\end{figure}

We now explore the phase diagram using both methods by 
first considering a case where correlated hopping dominates ($t'=0.95$). The
phase diagram in the ($1/V,\mu/V$) plane deduced from QMC and from the
SCA is summarized in Fig.~\ref{fig:pdV}, while simulation
results for $V=2.2$ ($1/V=0.45$) are shown in the left panel of Fig.~\ref{fig:SS_T0}.
Two features of this phase diagram are striking: First of all, the large $V$ (small $1/V$)
region is dominated by a very large supersolid phase that appears for densities above $1/2$. 
Secondly, this supersolid phase extends far below the critical value $V_c^{CB}=2.38$ ($1/V_c^{CB}=0.42$) for the development of 
CB order, down to $V_c^{SS}=1.74$ ($1/V_c^{SS}=0.57$). In other words, with correlated hopping, {\it a supersolid phase can exist 
without a neighboring solid phase} as $\mu$ is changed. For frustrated quantum antiferromagnets,
this implies that supersolid phases can show up even in the absence of
magnetization plateaus.
Note that correlated hopping appears to be crucial for this physics to be
realized. Single particle further-neighbour hopping has been recently shown
to induce supersolid phases near different kinds of solid phases~\cite{chen07}, but, as
far as we can tell, never to stabilize a supersolid phase without an
adjacent solid one. 

Upon increasing the chemical potential $\mu$, the SF-to-solid and SF-to-SS transitions are first order, while
the SS-to-SF and solid-to-SS transtions are second order, as illustrated in Fig.~\ref{fig:SS_T0}. 
Interestingly, the SCA and QMC agree on this point, as well as on the overall structure of the phase diagram. 
The SCA clearly overestimates the extent of the supersolid phase, but it correctly predicts that
it extends below $V_{CB}$. This is an important remark for more realistic models with possibly positive
correlated hopping amplitudes. In that case, QMC will not be possible due to the minus sign problem, but
the SCA can be expected to be qualitatively reliable.

\begin{figure}
    \begin{center}
        \includegraphics[width=.9\columnwidth]{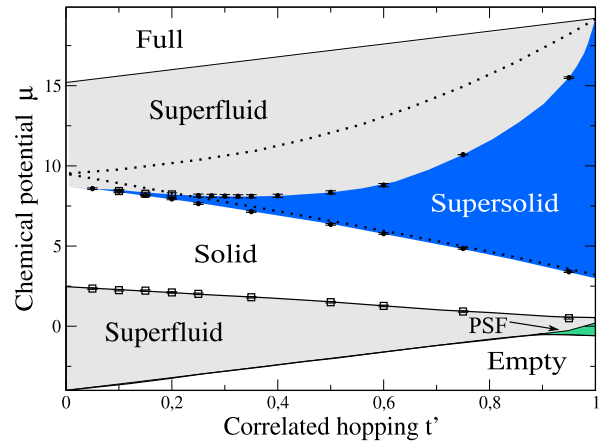}
    \end{center}
    \caption{(Color online) Zero-temperature phase diagram for $V=2.8$ as a function of the correlated hopping $t'$ versus
the chemical potential $\mu$.  Open squares (closed circles) denote first (second) order phase transitions deduced from QMC data. All the lines have been obtained by interpolating between numerical data. Thin dotted lines and the boundaries to the empty and full system are mean-field results.
}
    \label{fig:pd}
\end{figure}

Next, we investigate how this large supersolid phase evolves from the case
without correlated hopping ($t'=0$),
where there is no supersolid at all. We consider an intermediate value of the repulsion ($V=2.8$) and follow the
evolution of the $T=0$ phase diagram as a function of $t'$. The results in the plane ($t',\mu$)
are plotted in Fig.~\ref{fig:pd}. This phase diagram has been obtained with QMC simulations on lattices with up 
to $(24 \times 24)$ sites using temperatures $\beta=2 L$. The finite size effects are remarkably small, as illustrated in the right panel of Fig.~\ref{fig:SS_T0}, where the
results for $t'=0.75$ obtained on different clusters for $n$, $S(\pi,\pi)$ and $\rho_{\rm S}$ are plotted as a function of $\mu$.

A striking feature of this phase diagram is the strong asymmetry introduced by correlated hopping compared to the 
particle-hole symmetric situation when $t'=0$. When correlated hopping is introduced, the phase separation above 
the plateau is rapidly replaced by a supersolid phase, which grows continuously to extend over all the high density 
region when only correlated hopping is present. In contrast, phase separation persists below the solid phase 
for all values of $t'$.


The quantum phase transition between the solid and the SS falls into the conventional SF-insulator universality class,
as expected since the gapped excitations of the solid order are not expected to influence the nature of the quantum phase transition, and in agreement with the results recently reported for a spin model~\cite{laflo07}.
By contrast, the transition between the SS and the SF at zero temperature is first order below $t'\approx 0.25$, 
and seems to be continuous above. For the sizes available, the extracted critical exponents are consistent 
with the continuous transition being in the 3d Ising universality
class.
This might be a finite size effect though: The long-wavelength gapless excitations of the SF phase are
expected to change the universality class of this quantum phase transition~\cite{frey97}, and could therefore
give rise to a crossover phenomenon at large length scales.

Next we discuss the thermal transitions of the supersolid phase. In general one expects two melting transitions for a SS: a KT transition when the SF stiffness vanishes, and an another one when the solid order melts whose universality class depends on the type
of order~\cite{fisher74}. This question has already been addressed, first for hard-core bosons on the triangular lattice~\cite{bonin05}, 
then for a spin model with anisotropic exchange integrals.~\cite{laflo07} In both cases, two phase transitions have indeed been observed. In the case of the spin model, closer to the present case since the melting of the solid is in the Ising universality class, the KT transition has been found to lie always below the Ising transition, suggesting that the supersolid needs a solid phase to develop. In the present case, we expect the situation to be quite different since at zero temperature
a supersolid can exist without a solid. This is confirmed by our investigation of the two representative cases for which zero temperature data have been shown in Fig.~3: $t'=0.75$ and $V=2.8$ (Fig.~5a and Fig.~5b) and $t'=0.95$ and $V=2.2$ (Fig.~5c). 
For the first case, we indeed find two transition lines
which both smoothly go to $T=0$ upon approaching either the solid phase (KT) or the SF phase (Ising).
They cross in the middle of the SS phase, defining a region close to the solid where the KT transition is below the Ising one, as in Ref.~\onlinecite{laflo07}, but also a region close to the SF where the Ising transition is below the KT one. This is even more dramatic for the second case of Fig. 5c, where the melting of the Ising order occurs entirely inside the superfluid phase.

\begin{figure}
    \begin{center}
    \includegraphics[width=\linewidth]{fig5a}\vspace*{3mm}
    \includegraphics[width=\linewidth]{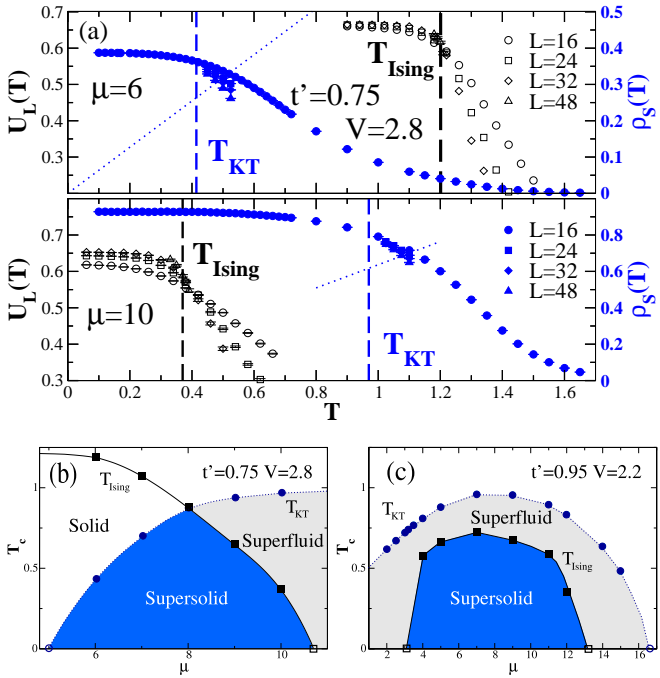}
    \end{center}
    \caption{(Color online) 
     (a) Thermal melting of the supersolid for $t'=0.75$ and $V=2.8$: Binder ratio (empty black symbols) and superfluid stiffness (filled blue symbols) as a function of temperature for $\mu=6$ (upper panel) and $\mu=10$ (lower panel). The location of the melting transitions are marked by vertical dashed lines.
    (b) Finite-temperature phase diagram as a function of the chemical potential for $t'=0.75$ and $V=2.8$ ($1/V=0.36$). 
    (c) Same as (b) for $t'=0.95$ and $V=2.2$ ($1/V=0.455$).
    In (b) and (c), lines are interpolations between numerical data, and
     error bars are smaller than the symbols. 
    }
    \label{SS_FiniteT}
\end{figure}

Finally, let us comment on the experimental implications of these results. Whenever correlated hopping 
dominates, one can reasonably expect to find supersolid phases, even in regions where no plateau 
has been detected. 
In SrCu$_2$(BO$_3$)$_2$, such a large domain exists between the 1/8 and 1/4
plateaux, in a field range accessible to NMR, a technique well suited to
detect lattice symmetry breaking. 
Even if experiments such as NMR, which are almost always done in steady field,
cannot be performed, our results firmly establish the presence 
of two phase transitions, the melting of the solid often taking place below
the KT transition if correlated hopping is present. 
Since the magnetization is expected to show only a very weak anomaly (if any), it would be very interesting to perform 
systematic specific heat measurements of frustrated dimer antiferromagnets
in high field and at very low temperature to try 
to detect ordering in region where the magnetization does not exhibit any plateau. It is our hope that
the present paper will stimulate such investigations.

\acknowledgments
We thank L.~Balents, N.~Laflorencie and S.~Wessel for stimulating discussions. 
The SSE simulations were done using a modified SSE code \cite{alet05} of the ALPS project~\cite{alps}. 
We acknowledge the Swiss National Funds and the MaNEP for financial support 
and the CSCS (Manno) for allocation of computing time.


\begin{thebibliography}{10}
\bibitem{wesse05}
S. Wessel and M. Troyer, Phys. Rev. Lett. {\bf 95}, 127205 (2005).
\bibitem{heida05}
D. Heidarian and K. Damle, Phys. Rev. Lett. {\bf 95}, 127206 (2005).
\bibitem{melko05}
R. Melko {\it et al.}, Phys. Rev. Lett. {\bf 95}, 127207 (2005).
\bibitem{momoi00}
T. Momoi and K. Totsuka, Phys. Rev. B {\bf 62}, 15067 (2000).
\bibitem{ng06}
 K.-K. Ng and T. K. Lee, Phys. Rev. Lett. {\bf 97}, 127204 (2006).
\bibitem{sengu07}
P. Sengupta and C. D. Batista, Phys. Rev. Lett. {\bf 98}, 227201 (2007).
\bibitem{laflo07}
N. Laflorencie and F. Mila, Phys. Rev. Lett.  {\bf 99}, 027202 (2007).
\bibitem{mila}
 F. Mila, Eur. Phys. J. B {\bf 6}, 201 (1998). 
\bibitem{batro00}
G.G. Batrouni and R.T. Scalettar, Phys. Rev. Lett. {\bf 84}, 1599 (2000).
\bibitem{momoi00_2}
T. Momoi and K. Totsuka, Phys. Rev. B {\bf 61}, 3231 (2000).
\bibitem{shinreview}
S. Miyahara and K. Ueda, J. Phys. Condens. Matter {\bf 15}, R327 (2003).
\bibitem{sandvik}
A.W. Sandvik and J. Kurkijarvi, Phys. Rev. B {\bf 43}, 5950 (1991);
O.F.~Syljuasen and A.W. Sandvik, Phys. Rev. E {\bf 66}, 046701 (2002).
\bibitem{schmi06}
K.P. Schmidt {\it et al.},
Phys. Rev. B {\bf 74}, 174508 (2006).
\bibitem{schmid02}
G. Schmid {\it et al.},
Phys. Rev. Lett. {\bf 88}, 167208 (2002).
\bibitem{scalet95}
R. T. Scalettar {\it et al.},
Phys. Rev. B {\bf 51}, 8467 (1995).
\bibitem{matsubara56} T. Matsubara and H. Matsuda, Prog. Theor. Phys. 
{\bf 16}, 569 (1956).
\bibitem{note} Due to this non-collinearity, the soft-core boson
model that describes the fluctuations around the classical ground
state is in general very different from the soft-core version of the original
hard-core boson model.
\bibitem{chen07} Y.C. Chen, R. G. Melko, S. Wessel, Y.-J. Kao, unpublished
  (arXiv:0708.1807);
K.-K. Ng, Y.C. Chen, unpublished (arXiv:0710.3463).
\bibitem{frey97}
E. Frey and L. Balents, Phys. Rev. B {\bf 55}, 1050 (1997).
\bibitem{fisher74}
M.E. Fisher and D.R. Nelson, Phys. Rev. Lett. {\bf 32}, 1350 (1974).
\bibitem{bonin05}
M. Boninsegni and N. Prokof'ev, Phys. Rev. Lett. {\bf 95}, 237204 (2005).
\bibitem{alet05}
F. Alet {\it et al.},
Phys. Rev. E {\bf 71}, 036706 (2005). 
\bibitem{alps}
A. F. Albuquerque {\it et al.}, J. Magn. Magn. Mater. {\bf 310}, 1187 (2007);
M. Troyer {\it et al.}, Lecture Notes in Computer Science, {\bf 1505}, 191 (1998). 
\end{thebibliography}
\end{document}